\documentclass[pra,showpacs,twocolumn]{revtex4}
\usepackage{amsmath}
\usepackage{amssymb}
\usepackage{graphicx}
\usepackage{epsfig}

\begin{document}

\title{Information-Theoretic Measure of Genuine Multi-Qubit Entanglement }
\author{Jian-Ming Cai$^{\text{1}}$, Zheng-Wei Zhou$^{\text{1}}$}

\email{zwzhou@ustc.edu.cn}

\author{ Xing-Xiang Zhou$^{\text{2}}$, and
Guang-Can Guo$^{\text{1}}$}

\address{$^{\text{1}}$Laboratory of Quantum Information,
University of Science and \\ Technology of China, Hefei, Anhui
230026, P. R. China\\ $^{\text{2}}$Physics Department,
Pennsylvania State University, University Park, Pennsylvania
16802, USA}

\begin{abstract}
We consider pure quantum states of $N$ qubits and study the
genuine $N-$qubit entanglement that is shared among all the $N$
qubits. We introduce an information-theoretic measure of
\textit{genuine $N$-qubit entanglement} based on bipartite
partitions. When $N$ is an even number, this measure is presented
in a simple formula, which depends only on the purities of the
partially reduced density matrices. It can be easily computed
theoretically and measured experimentally. When $N$ is an odd
number, the measure can also be obtained in principle.

\end{abstract}

\maketitle

The nature of quantum entanglement is a fascinating topic in
quantum mechanics since the famous Einstein-Podolsky-Rosen paper
\cite{EPR} in 1935. Recently, much interest has been focused on
entanglement in quantum systems containing a large number of
particles. On one hand, multipartite entanglement is valuable
physical resource in large-scale quantum information processing
\cite{Raussendorf,Pan}. On the other hand, multipartite
entanglement seems to play an important role in condensed matter
physics \cite{Bulk Property}, such as quantum phase transitions
(QPT) \cite {Entanglement and Phase Transitions,Osborne032110} and
high temperature superconductivity \cite{HSC}. Therefore, how to
characterize and quantify multipartite entanglement remains one of
the central issues in quantum information theory.

In the present literature, there exist very few measures of
multipartite entanglement with a clear physical meaning
\cite{Wong,Hyperdeterminants,Mayer,GE,Gerardo,Mintert1,Barnum,You}.
Because of this, most research in quantum entanglement and QPT
focused on bipartite entanglement, for which there have been
several well defined measures
\cite{Plenio,EOF,ConC,negativity,REE,Eoa,Mintert2}. However,
bipartite entanglement can not characterize the global quantum
correlations among all parties in a multiparticle system. Since
the correlation length diverges at the critical points,
multipartite entanglement plays an essential role in QPT. Though
localizable entanglement (LE) \cite{Localizable entanglement} can
be used to describe long-range quantum correlations, its
determination is a formidable task for generic pure states.
Therefore, computable measure of multipartite entanglement with
clear physical meanings are highly desired \cite{Lunkes,SLOCC}.

In this paper, we define a new measure of genuine $N-$qubit
entanglement based on different bipartite partitions of the qubits
and existing measures for mutual information. The central idea is
that, through bipartite partitions, we can get information about
the genuine multi-qubit entanglement. Our measure is a polynomial
SLOCC (stochastic local operations and classical communication)
invariant \cite{Leifer} and is unchanged under permutations of
qubits. When $N$ is an even number, we derive a simple formula for
this measure, which is determined by the purity of partially
reduced density matrices only. It can be computed through
experimentally observable quantities \cite{GE,Mintert3}.
Therefore, it is easy to obtain not only theoretically but also
experimentally. For $N=4$, we show that this measure satisfies all
the necessary conditions required for a natural entanglement
measure \cite{REE} exactly. When $N$ is an odd number, the measure
for genuine $N$-qubit entanglement is defined on the basis of the
measure for even number of qubits. This measure will definitely
extend the research in the field of multipartite entanglement and
condensed matter systems.

\textit{Bipartite partition and genuine multi-qubit entanglement}
For multi-qubit pure states, there exist local information, and
nonlocal information which is related to quantum correlations
\cite{Horodecki}. In a closed two- and three- qubit system, i.e.
the system state is pure and its evolution is unitary, we have
shown that \cite{ICL} entanglement is relevant to some kind of
nonlocal information which contributes to the conserved total
information. Consider a pure state $|\psi\rangle$ of $N$ qubits,
labelled as $1,2,\cdots ,N$, generally we can write
$|\psi\rangle=|\psi_{1}\rangle \otimes \cdots \otimes
|\psi_{M}\rangle$ \cite{SLOCC}, where $|\psi_{m}\rangle$ are
non-product pure states, $m=1,2,\cdots, M$, and the qubits of
different $|\psi_{m}\rangle$ have no intersection. If $M = 1$,
$|\psi\rangle$ itself is a non-product pure state, otherwise
$|\psi\rangle$ is a product pure state. A bipartite partition
$\mathcal{P}$ will divide the qubits of $|\psi_{m}\rangle$ into
two subsets $\mathcal{A}_{m}$ and $\mathcal{B}_{m}$, then all the
$N$ qubits are divided into $\mathcal{A}=\cup
_{m=1}^{M}\mathcal{A}_{m}$ and $\mathcal{B}=\cup
_{m=1}^{M}\mathcal{B}_{m}$, e.g. see Fig.1(a). In this paper, we
use the linear entropy \cite{linear entropy}, then the mutual
information between $\mathcal{A}_{m}$ and $\mathcal{B}_{m}$ is
$I_{\mathcal{A}_{m}\mathcal{B}_{m}}=S_{\mathcal{A}_{m}}+S_{\mathcal{B}_{m}}-S_{\mathcal{A}_{m}\mathcal{B}_{m}}$,
where $S_{Y}=1-Tr\rho _{Y}^{2}$,
$Y=\mathcal{A}_{m},\mathcal{B}_{m},\mathcal{A}_{m}\mathcal{B}_{m}$.
Since $|\psi_{m}\rangle$ is pure, we can write
$I_{\mathcal{A}_{m}\mathcal{B}_{m}}=2(1-Tr\rho
_{\mathcal{A}_{m}}^{2})=2(1-Tr\rho _{\mathcal{B}_{m}}^{2})$, where
$\rho _{\mathcal{A}_{m}}$ and $\rho _{\mathcal{B}_{m}}$ are the
reduced density matrices. The bipartite nonlocal information
between $\mathcal{A}$ and $\mathcal{B}$, denoted as
$S_{\mathcal{A}|\mathcal{B}}$, is defined as the sum of mutual
information between $\mathcal{A}_{m}$ and $\mathcal{B}_{m}$,
\begin{equation}
S_{\mathcal{A}|\mathcal{B}}:=\sum \limits_{m=1}^{M}
I_{\mathcal{A}_{m}\mathcal{B}_{m}}
\end{equation}
If $\mathcal{A}_{m}$ or $\mathcal{B}_{m}$ is empty we set
$I_{\mathcal{A}_{m}\mathcal{B}_{m}}=0$.

\begin{figure}[htb]
\epsfig{file=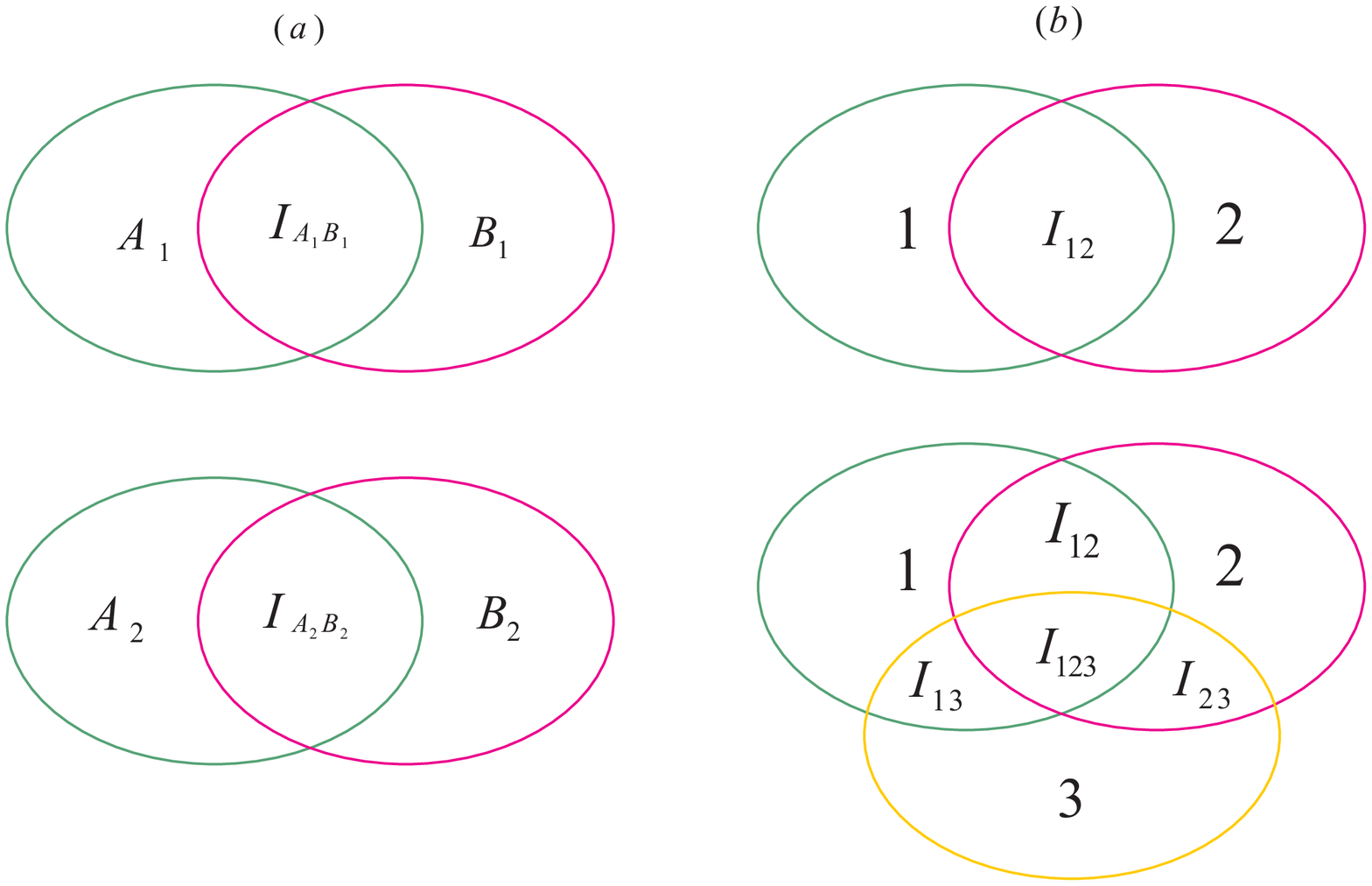,width=6cm,}
\caption{Information diagram for (a) product pure states
$|\psi\rangle_{A_{1}B_{1}}\otimes|\psi\rangle_{A_{2}B_{2}}$, the
bipartite nonlocal information (\textit{overlap}) between
$A=A_{1}\cup A_{2}$ and $B=B_{1}\cup B_{2}$ is
$S_{A|B}=I_{A_{1}B_{1}}+I_{A_{2}B_{2}}$; (b) two-qubit (top) and
three-qubit (bottom) pure states.}
\end{figure}

The information diagram for two- and three-qubit pure states is
depicted in Fig.1(b). Each qubit is represented by one circle. The
overlap of $k$ different circles, indexed as $i_{1}, i_{2},
\cdots, i_{k}$, represents the nonlocal information $I_{i_{1}
i_{2} \cdots i_{k}}$ that is \textit{shared among all these $k$
qubits}. If we adopt some appropriate measure of information, i.e.
linear entropy \cite{ICL,linear entropy}, the nonlocal information
is directly relevant to genuine $k-$qubit entanglement. For
example, for two-qubit pure states, $I_{12}=\tau_{12}$, where
$\tau_{12}$ is the square of concurrence. For three-qubit pure
states, the bipartite nonlocal information between $1$ and $23$
can be written as $S_{1|23}=I_{12}+I_{13}+I_{123}$ with
$I_{12}=\tau_{12}$, $I_{13}=\tau_{13}$ and $I_{123}=\tau_{123}$,
where 3-tangle $\tau_{123}$ has been shown to be a well-defined
measure of genuine three-qubit entanglement \cite{CKW}. For pure
states of $N>3$ qubits, we generalize the above viewpoint of
multi-qubit entanglement, assisted by the information diagram, to
quantify genuine $N$-qubit entanglement by the nonlocal
information shared by all the $N$ qubits.

\textit{Observation: For a pure state of $N$ qubits and a
partition $\mathcal{A}|\mathcal{B}$, the bipartite nonlocal
information between $\mathcal{A}$ and $\mathcal{B}$ is contributed
by different levels of nonlocal information $S_{
\mathcal{A}|\mathcal{B}}=\sum\limits_{k=2}^{N}I_{i_{1}i_{2}\cdots
i_{k}}$, where $i_{1}, i_{2}, \cdots, i_{k}$ are not in the same
set $\mathcal{A}$ or $\mathcal{B}$, and $I_{i_{1}i_{2}\cdots
i_{k}}$ is some appropriate measure of nonlocal information.}

Based on the above observation, we propose an
information-theoretic measure of genuine $N-$qubit entanglement
through bipartite partitions. We utilize this measure to explore
the genuine multi-qubit entanglement in spin systems. Our results
also suggest that the above observation is reasonable.

\textit{Genuine four-qubit entanglement} For a pure state of $N$
qubits, where $N\in even$, there are two different classes of
bipartite partitions $\mathcal{P}_{\mathcal{I}}$ and
$\mathcal{P}_{ \mathcal{II}}$. For any partition
$\mathcal{P}=\mathcal{A}|\mathcal{B}$, we denote the number of
qubits contained in $\mathcal{A}$ and $\mathcal{B}$ as
$|\mathcal{A}|$, $| \mathcal{B}|$. If $|\mathcal{A}|,|
\mathcal{B}| \in odd$, $\mathcal{P}\in \mathcal{P}_{\mathcal{I}}$,
and if $|\mathcal{A}|,| \mathcal{B}| \in even$, $\mathcal{P}\in
\mathcal{P}_{\mathcal{II}}$. When $N=4$,
$\mathcal{P}_{\mathcal{I}}=\{1|234, 2|134, 3|124, 4|123\}$ and
$\mathcal{P}_{\mathcal{II}}=\{12|34, 13|24, 14|23\}$. For
partition $1|234$, the bipartite nonlocal information is denoted
as $S_{1|234}=I_{12}+ I_{13}+I_{14}+ I_{123}+I_{124}+I_{134}+
I_{1234}$. Similarly, we can get the bipartite nonlocal
information for the other seven partitions, denoted as $S
_{2|134}$, $S_{3|124}$, $S_{4|123}$, $S_{12|34}$, $S_{13|24}$ and
$S_{14|23}$. It can be seen that $S_{\mathcal{I}
}=\sum\limits_{\mathcal{P} \in \mathcal{P}_{ \mathcal{I}}}S_{
\mathcal{P}}=2I_{12}+2I _{13}+2I_{14}+ 2 I_{23}+2I_{24}+2I _{34}+
3I_{123}+3 I_{124}+3I_{134}+3I _{234}+ 4I_{1234}$. In the same
way, we can get $S_{\mathcal{II}}=\sum\limits_{\mathcal{P} \in
\mathcal{P}_{ \mathcal{II}}}S_{ \mathcal{P}}=S_{\mathcal{I}
}-I_{1234}$. Therefore, genuine four-qubit entanglement
$\mathcal{E}_{1234}$ can be naturally measured by the nonlocal
information $I_{1234}$, i.e. the difference between $S_{
\mathcal{I}}$ and $ S_{\mathcal{II}}$
\begin{equation}
\mathcal{E}_{1234}=S_{\mathcal{I}}-S_{\mathcal{II}}
\end{equation}

The above definition of the measure for genuine four-qubit
entanglement must satisfy the following conditions in order to be
a natural entanglement measure for pure states \cite{REE}. $(1)$
$\mathcal{E}_{1234}$ is invariant under local unitary operations.
$(2)$ $\mathcal{E}_{1234}\geq 0$ for all pure states. $(3)$
$\mathcal{E}_{1234}$ is an entanglement monotone, \textit{i.e.} ,
$\mathcal{E}_{1234}$ does not increase on average under local
quantum operations assisted with classical communication (LOCC).

It is obvious that the mutual information
$I_{\mathcal{A}_{m}\mathcal{B}_{m}}$ is determined by the
eigenvalues of the partially reduced density matrices. Therefore,
\textit{$\mathcal{E}_{1234}$} is invariant under local unitary
operations. Moreover, according to the above definitions in
Eqs.(1-2), it is easy to verify that $\mathcal{E}_{1234}=0$ for
product pure states.

In order to prove that $\mathcal{E}_{1234}$ satisfies the other
two conditions, we first investigate how $\mathcal{E}_{1234}$ will
change under determinant 1 SLOCC operations. Determinant 1 SLOCC
operations \cite {SLOCC, W states} are local operations which
transform a pure state $ |\psi\rangle$ to
$|\psi^{\prime}\rangle=A_{1}\otimes \cdots \otimes
A_{n}|\psi\rangle/Q$ with $Q^2=Tr(A_{1}\otimes \cdots \otimes
A_{n}|\psi\rangle\langle\psi|A^{\dag}_{1}\otimes \cdots \otimes
A^{\dag}_{n}) $, where $A_{i}\in SL(2,C)$ is an operator on the
$ith$ qubit. Without loss of generality, we assume that $
|\psi\rangle$ is a non-product pure state, and the determinant 1
SLOCC operation is only performed on the $1st$ qubit. The
determinant 1 SLOCC operation $A_{1}$ can be written as $U\cdot
diag[d,1/d] \cdot V$, where $U$, $V$ $\in SU(2)$ and $d$ is a
positive real number. Since the mutual information is invariant
under local unitary operations, we do not have to consider the
unitary operations $U$ and $V$. Let us write the four-qubit state
$|\psi\rangle=\sqrt{p_{0}}|0\rangle|\varphi_{0}\rangle+\sqrt{p_{1}}
|1\rangle|\varphi_{1}\rangle$, where $p_{0}+p_{1}=1$,
$|\varphi_{0}\rangle$ and $|\varphi_{1}\rangle$ are pure states of
qubits $2,3,4$. After the operation $A_{1}$, $
|\psi^{\prime}\rangle=\frac{A_{1}}{Q}|\psi\rangle=\frac{\sqrt{
p_{0}}d}{Q}|0\rangle|\varphi_{0}\rangle+\frac{\sqrt{p_{1}}}{Q d}
|1\rangle|\varphi_{1}\rangle$, where
$Q=(p_{0}d^2+p_{1}/d^2)^{1/2}$. Since
$\rho_{234}^{\prime}=(p_{0}d^2|
\varphi_{0}\rangle\langle\varphi_{0}|+p_{1}/d^2|\varphi_{1}\rangle\langle
\varphi_{1}|)/Q^2$, $Tr(\rho_{1}^{\prime})^2=Tr(\rho_{234}^{
\prime})^2=(p_{0}^{2} d^4+p_{1}^{2}/d^4+f_{1})/Q^4$, where $
f_{1}=2p_{0}p_{1}Tr(|\varphi_{0}\rangle\langle\varphi_{0}|\varphi_{1}\rangle
\langle\varphi_{1}|)$ is independent on the value of $d$. Then we
can get the bipartite nonlocal information for partition $1|234$
as $S_{1|234}^{\prime}=2-2(p_{0}^{2}
d^4+p_{1}^{2}/d^4+f_{1})/Q^4$. In the same way, after
straightforward calculation, we can get the other bipartite
nonlocal information $S_{i|\tilde{i}}^{\prime}=2-2[p_{0}^{2} d^4
Tr\rho^2_{i}(0)+p_{1}^{2}/d^4 Tr\rho^2_{i}(1)+f_{i}]/Q^4$ and
$S_{ij|\tilde{ij}}^{\prime}=2-2[p_{0}^{2} d^4
Tr\rho^2_{ij}(0)+p_{1}^{2}/d^4 Tr\rho^2_{ij}(1)+f_{ij}]/Q^4$,
where $\tilde{i}$, $\tilde{ij}$ represent the qubits other than
$i$ and $i,j$, $\rho_{i}(k)$ ($i=2,3,4$) are the reduce density
matrices of the $ith$ qubit and $\rho_{ij}(k)$ ($ij=23,24,34$) are
the reduce density matrix of the $ith$ and $jth$ qubit from $
|\varphi_{k} \rangle$ ($k=0,1$). $f_{i}$ and $f_{ij}$ are all
independent on the value of $d$. We note that $Tr\rho^2_{2}(k)=Tr%
\rho^2_{34}(k)$, $Tr\rho^2_{3}(k)=Tr\rho^2_{24}(k)$ and $Tr\rho^2_{4}(k)=Tr%
\rho^2_{23}(k)$. It can been seen that $\mathcal{E}
_{1234}(|\psi^{\prime}\rangle\langle\psi^{
\prime}|)=2(2p_{0}p_{1}-f_{1}-f_{2}-f_{3}-f_{4}+f_{23}+f_{24}+f_{34})/Q^4$,
i.e.
\begin{equation}
\mathcal{E}_{1234}(|\psi^{\prime}\rangle\langle\psi^{\prime}|)=\mathcal{E}
_{1234}(|\psi\rangle\langle\psi|)/Q^4
\end{equation}
If $ |\psi\rangle$ is a product pure state, $\mathcal{E} _{1234}$
is always zero, i.e. it satisfies the above Eq.(3) too. Therefore,
if we take into account the normalization factor, $\mathcal{E}
_{1234}$ is SLOCC invariant \cite{Normal forms}.

We now prove that $\mathcal{E}_{1234}$ satisfies the above
condition (2), i.e. $\mathcal{E}_{1234}\geq 0$. It has been shown
that \cite{Normal forms} all pure multipartite states can be
transformed into a normal form, all local density operators of
which are proportional to the identity, by the determinant 1 SLOCC
operations. A generic pure state of four qubits can always be
transformed to the normal form state \cite{Verstraete2002} $
G_{abcd}=\frac{a+d}{2}(|0000\rangle+|1111\rangle)+\frac{a-d}{2}
(|0011\rangle+|1100\rangle)+\frac{b+c}{2}(|0101\rangle+|1010\rangle)+\frac{
b-c}{2}(|0110\rangle+|1001 \rangle)$, where $a, b, c ,d$ are
complex parameters with nonnegative real part. We denote
$A=(a+d)/2, B=(b+c)/2, C=(a-d)/2, D=(b-c)/2$. If $G_{abcd}$ is a
product state, $\mathcal{E}_{1234}=0$. Again we assume $G_{abcd}$
is a non-product state in the following. The bipartite nonlocal
information of different partitions is give by $
S_{1|234}=S_{2|134}=S_{3|124}=S_{4|123}=1$. $S_{12|34}$,
$S_{13|24}$ and $S_{14|23}$
can be obtained from $Tr\rho^2_{12}=\mathcal{T}(A,B,C,D)$, $Tr\rho^2_{13}=%
\mathcal{T}(A,C,B,D)$ and $Tr\rho^2_{14}=\mathcal{T}(A,B,D,C)$, where $%
\mathcal{T}(x_{1},x_{2},x_{3},x_{4})=2
\{(|x_{1}|^2+|x_{3}|^2)^2+(|x_{2}|^2+|x_{4}|^2)^2+|x_{1}x_{3}^{*}+x_{1}^{*}x_{3}|^2+|x_{2}x_{4}^{*}+x_{2}^{*}x_{4}|^2\}/
\mathcal{M}^2 $ with
$\mathcal{M}=2(|x_{1}|^2+|x_{2}|^2+|x_{3}|^2+|x_{4}|^2)$. We first
prove the following inequality $
\sum\limits_{i=1}^{4}|x_{i}|^{4}+\sum\limits _{\substack{ i=1  \\ j>i}}%
^{4}|x_{i}x_{j}^{*}+x_{i}^{*}x_{j}|^{2}\geq 2\sum\limits
_{\substack{ i=1\\ j>i}}^{4}|x_{i}|^2|x_{j}|^2$. Without loss of
generality, we could assume that $x_{1}=|x_{1}| \in \mathcal{R}$
and $x_{i}=|x_{i}|\exp{(i\phi_{i}/2)}$ with $i=2,3,4$. The inequality is equivalent to $\sum%
\limits_{i=1}^{4}|x_{i}|^{4}+2(\sum\limits_{i=2}^{4}|x_{1}|^2|x_{i}|^2\cos{%
\phi_{i}} +|x_{2}|^2|x_{3}|^2\cos{(\phi_{2}-\phi_{3})}+|x_{2}|^2|x_{4}|^2%
\cos
{(\phi_{2}-\phi_{4})}+|x_{3}|^2|x_{4}|^2\cos{(\phi_{3}-\phi_{4})})
=(|x_{1}|^2+\sum\limits_{i=2}^{4}|x_{i}|^2\cos{\ \phi_{i}}%
)^2+(\sum\limits_{i=2}^{4}|x_{i}|^2\sin{\ \phi_{i}})^2 \geq 0$.
According to this inequality, is can be easily verified that
\begin{equation}
\mathcal{E}_{1234}(G_{abcd})\geq 0
\end{equation}
Together with Eq.(3), we obtain that
$\mathcal{E}_{1234}(|\psi\rangle \langle\psi|)\geq 0$ for any
four-qubit pure state.

The remain condition is that $\mathcal{E}_{1234}$ should be an
entanglement monotone. Note that any local protocol can be
decomposed into local positive-operator valued measure (POVMs)
that can be implemented by a sequence of two-outcome POVMs
performed by one party on the system \cite{W states}. Without loss
of generality, we only need to consider the general two-outcome
POVMs performed on the $1st$ qubit $\{M_{1}, M_{2}\}$, such that $
M_{1}^{\dag}M_{1}+M_{2}^{\dag}M_{2}=\mathcal{I}$. Using the
singular value decomposition, we can write $ M_{1}=U_{1}\cdot
diag\{a,b\}\cdot V, M_{2}=U_{2}\cdot diag\{\sqrt{1-a^2},
\sqrt{1-b^2}\}\cdot V$, where $U_{1}$, $U_{2}$ and $V$ are unitary
matrices. We denote that $M^{\prime}_{1}=M_{1}/(ab)^{1/2},
M^{\prime}_{2}=M_{2}/\{(1-a^2)(1-b^2)\}^{1/4}$. Note that $\det
M^{\prime}_{1}=\det M^{\prime}_{2}=1$. According to Eq.(3), we get
$ \mathcal{E}_{1234}(|\psi_{1}\rangle\langle\psi_{1}|)=\frac{a^2
b^2}{ Q^4_{1} }\mathcal{E}_{1234}(|\psi\rangle\langle\psi|)$,
$\mathcal{E}_{1234}(|\psi_{2}\rangle\langle\psi_{2}|)=\frac{(1-a^2)(1-b^2)}{
Q^4_{2}}\mathcal{E}_{1234}(|\psi\rangle\langle\psi|)$, where
$|\psi_{i}\rangle=M_{i}|\psi\rangle/Q_{i}$ and $Q_{i}$ are
normalization factors. After simple algebra calculation and using
the fact that the arithmetic mean always exceeds the geometric
mean \cite{Wong,W states,Normal forms}, we can see that
\begin{equation}
\nonumber
Q^2_{1}\mathcal{E}_{1234}(|\psi_{1}\rangle\langle\psi_{1}|)+Q^2_{2}\mathcal{E}
_{1234}(|\psi_{2}\rangle\langle\psi_{2}|)\leq
\mathcal{E}_{1234}(|\psi \rangle\langle\psi|)
\end{equation}
is fulfilled, i.e. $\mathcal{E}_{1234}$ is an entanglement
monotone. Therefore, it satisfies all the necessary conditions for
a natural entanglement measure.

\textit{General even number qubits} The above discussions can be
extended to the situation when $N$ is a generic even number. We
denote the set of all genuine $k-$qubit entanglement that
contributes to the bipartite nonlocal information of partition
$\mathcal{P}\in \mathcal{P_{I}}$ and $\mathcal{P}\in
\mathcal{P_{II}}$ as $ \mathcal{EC}_{\mathcal{I}}^{k}$,
$\mathcal{EC}_{\mathcal{II}}^{k}$ respectively. If $k=N$, the
contribution is just genuine $N-$qubit nonlocal information
$I_{12\cdots N}$, and
$|\mathcal{EC}_{\mathcal{I}}^{N}|-|\mathcal{EC}_{\mathcal{II}}^{N}|=1$.
This can be seen from the polynomial expansion of
$(1-x)^{N}|_{x=1}=(2|\mathcal{EC
}_{\mathcal{II}}^{N}|+2)-2|\mathcal{EC}_{\mathcal{I}}^{N}|=0$. For
$2\leq k<N$, if the nonlocal information $I_{a_{1}a_{2}\cdots
a_{l}b_{1}b_{2}\cdots b_{k-l}}$ ($a_{1}, a_{2}, \cdots, a_{l}\in
\mathcal{A}$, $b_{1}, b_{2}, \cdots, b_{k-l}\in \mathcal{B}$)
contributes to the bipartite nonlocal information of some
partition $\mathcal{P}=\mathcal{A}|\mathcal{B}\in
\mathcal{P_{I}}$, there must exist one maximum index, denoted as
$\mathcal{X}$, which does not
belong to $\{a_{1},a_{2},\cdots ,a_{l},b_{1},b_{2},\cdots ,b_{k-l}\}$. If $%
\mathcal{X}\in \mathcal{A}$ or $\mathcal{B}$, then we can
construct a bipartite partition $ \mathcal{P^{\prime
}}=\mathcal{A}-\{\mathcal{X}\}|\mathcal{B}+\{\mathcal{X} \}$ or
$\mathcal{A}+\{\mathcal{X}\}|\mathcal{B}-\{\mathcal{X} \}$ $\in
\mathcal{P_{II}}$. The nonlocal information $I_{a_{1}a_{2}\cdots
a_{l}b_{1}b_{2}\cdots b_{k-l}}$ will also contributes to the
bipartite nonlocal information of partition $ \mathcal{P^{\prime
}}$. According to this one to one homologous relation, we can
obtain that
$\sum\limits_{k=2}^{N}(\mathcal{EC}_{\mathcal{I}}^{k}-\mathcal{EC}_{\mathcal{
II}}^{k})=I_{12\cdots N}$. Therefore, a measure for genuine
$N-$qubit entanglement, $N\in even $, of pure states can be
defined naturally as follows
\begin{equation}
\mathcal{E}_{12\cdots N}=\sum\limits_{\mathcal{P}\in
\mathcal{P_{I}}} S_{\mathcal{P}}-\sum\limits_{\mathcal{P}\in
\mathcal{P_{II}}} S_{\mathcal{P}}
\end{equation}
where $S_{\mathcal{P}}$ is the bipartite nonlocal information for
partition $\mathcal{P}$.

$\mathcal{E}_{12\cdots N}$ is SLOCC invariant, and is unchanged
under permutations of qubits, i.e., it represents a collective
property of all the $N$ qubits. Our measure can surely distinguish
two different kinds of multi-qubit entangled states, general
$N-$qubit GHZ states $|GHZ\rangle
_{N}=\frac{1}{\sqrt{2}}(|00\cdots 0\rangle +|11\cdots 1\rangle)$
and W states $|W\rangle _{N}=\frac{1}{\sqrt{N}}(|00\cdots 01
\rangle +|00\cdots 010\rangle +\cdots +|10\cdots 0\rangle )$, for
$\mathcal{E}_{12\cdots N}(|GHZ\rangle _{N})=1$ and
$\mathcal{E}_{12\cdots N}(|W\rangle _{N})=0$. Although we have not
proved that $\mathcal{E}_{12\cdots N}\geq 0$ for $N > 4$, which is
actually related to the intricate compatibility problem of
multipartite pure states \cite{Compatible}, we have calculated
numerically $\mathcal{E}_{12\cdots N}$ for more than $10^5$
arbitrarily chosen pure states of six and eight qubits. The
numerical results suggest strongly that $\mathcal{E}_{12\cdots
N}\geq 0$. The calculation of $ \mathcal{E}_{12\cdots N}$ is very
straightforward. It can indeed be determined from observable
quantities \cite{GE,Mintert3}, which can be conveniently measured
in experiments.

\textit{General odd number qubits} The above results are not
applicable to the situation of odd number qubits
straightforwardly. However, it is easy to verify that the genuine
$N-$qubit entanglement, $N\in odd$, can be characterized by the
bipartite nonlocal information of different partitions together
with genuine $(N-1)-$qubit entanglement, where $N-1$ is an even
number. Therefore, the measure can also be obtained based on our
idea in principle.

Based on the measure for pure states of $N$ qubits, the measure
for $N-$qubit mixed states is defined by the convex roof extension
of pure-state measure according to the standard entanglement
theory \cite{Plenio}, i.e.
\begin{equation}
\mathcal{E}_{12\cdots
N}(\rho)=\min\sum\limits_{k}p_{k}\mathcal{E}_{12\cdots
N}(|\psi_{k}\rangle\langle\psi_{k}|)
\end{equation}
where min runs through all possible decompositions of $\rho$ into
pure states, \textit{i.e.},
$\rho=\sum\limits_{k}p_{k}|\psi_{k}\rangle\langle \psi_{k}|$.

\begin{figure}[htb]
\epsfig{file=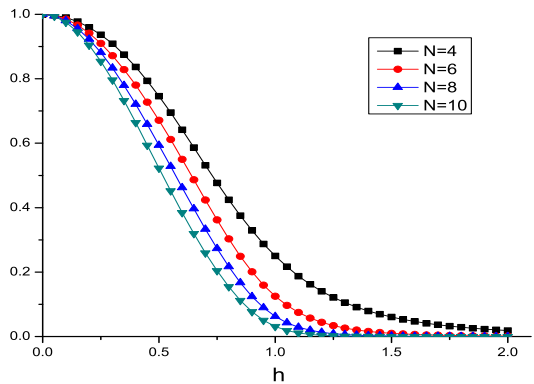,width=7cm,height=5cm,}
\caption{(Color online) Genuine $N-$qubit entanglement $\mathcal{
E }_{12\cdots N}$ for the ground state of the finite transverse
field Ising model with $N=4,6,8,10$.}
\end{figure}

\textit{Genuine multi-qubit entanglement in spin systems} Our
measure for genuine multi-qubit entanglement will extend the
research of the relation between entanglement and quantum phase
transitions. Given a quantum system with $N$ spins, one can
compute the genuine $N$-qubit entanglement of the ground state for
different even number $N$. In addition, the translational
invariant property together with other symmetries of the system
Hamiltonian will simplify the calculation of $\mathcal{ E
}_{12\cdots N}$ significantly. As an illustration, we consider the
ground state of the transverse field Ising Hamiltonian
$\mathcal{H}=-\sum\limits_{i=1}^{N}\sigma_{i}^{x}\sigma_{i+1}^{x}-h\sum\limits_{i=1}^{N}\sigma_{i}^{z}$.
We plot the behavior of $\mathcal{ E }_{12\cdots N}$ for different
system size $N=4,6,8,10$ (see Fig. 2). Using the standard
finite-size scaling theory, it can be seen that at the quantum
critical point $h_{c}=1$, genuine $N-$qubit entanglement changes
drastically. Compared to the results in Refs. \cite{Localizable
entanglement}, the behavior of $\mathcal{ E }_{12\cdots N}$ is
very similar to LE, i.e. it can capture the feature of long-range
quantum correlations.

In conclusion, we have introduced an information-theoretic measure
of genuine multi-qubit entanglement, which is the collective
property of the whole state. For pure states of an even number of
qubits, this measure is easily computable, and are dependent on
observable quantities. Therefore the measurement in experiments is
convenient. For the pure states of an odd number of qubits, the
measure is defined based on the results for the states of an even
number qubits. Finally, we demonstrated the usefulness of our
measure in spin systems. Further generalizations and application
will be presented in our future work. Our results will help gain
important insight into the structure and nature of multipartite
entanglement, and enlighten the research of quantum entanglement
in condensed matter physics.

\textbf{Acknowledgment} We gratefully acknowledge valuable
discussions with Lu-Ming Duan, Yong Hu, Yong-Jian Han and Xiang-Fa
Zhou. This work was funded by National Fundamental Research
Program (2001CB309300), the Innovation funds from Chinese Academy
of Sciences, NCET-04-0587, and National Natural Science Foundation
of China (Grant No. 60121503, 10574126).

\end{document}